\documentclass[aps,twocolumn,prd,nofootinbib,superscriptaddress]{revtex4-2}
\usepackage[utf8]{inputenc} % encoding
\usepackage{graphicx}       % graphics
\usepackage{dcolumn}    % for decimal-aligned columns in tables.
\usepackage[colorlinks=true]{hyperref}
\usepackage{amsmath}    % math formatting
\usepackage{bm}         % bold math symbols
\usepackage{xcolor}     % for colored text
\usepackage{lipsum}     % for placeholders
\usepackage{orcidlink}  % to add Orcid link
\usepackage{macros}

\def\invMpc{{\rm Mpc}^{-1}}
\def\bs{\boldsymbol}
\newcommand{\code}[1]{\texttt{#1}}
\newcommand{\fede}{f_{\rm EDE}}

\begin{document}

\title{Early Dark Energy Effects on the 21cm Signal}
\author{Tal Adi\,\orcidlink{0000-0002-5763-9353}}
\email{talabadi@post.bgu.ac.il}
\affiliation{Department of Physics, Ben-Gurion University of the Negev, Be’er Sheva 84105, Israel}

\author{Jordan Flitter\,\orcidlink{0000-0001-8092-8228}} 
\email{jordanf@post.bgu.ac.il}
\affiliation{Department of Physics, Ben-Gurion University of the Negev, Be’er Sheva 84105, Israel}

\author{Ely D. Kovetz\,\orcidlink{0000-0001-9256-1144}}
\email{kovetz@bgu.ac.il}
\affiliation{Department of Physics, Ben-Gurion University of the Negev, Be’er Sheva 84105, Israel}

\begin{abstract}

Early dark energy (EDE) is one of the leading models proposed to resolve the perplexing Hubble tension. Despite extensive scrutiny and testing against various observables, conclusive constraints remain elusive as we await new data. In this paper, we study the impact of EDE on the 21cm signal, a powerful probe of cosmic dawn, and the epoch of reionization. First, we examine the signatures of the shift in cosmological parameters and the new EDE parameters on the evolution of the 21cm signal compared to $\Lambda$CDM. We then focus on the implications of these signatures for upcoming radio interferometer telescopes, such as the Hydrogen Epoch of Reionization Array (HERA), and their ability to differentiate between EDE and $\Lambda$CDM. Finally, we forecast HERA's sensitivity to the fractional energy density of EDE, $\fede$, assuming a fiducial EDE model. We find significant modifications to the 21cm signal due to the presence of EDE. Furthermore, our analysis suggests that HERA, operating in its designed configuration, is poised to differentiate between the models and be sensitive to $\fede$ within $2\sigma$ after $\mathcal{O}(100)$ days of observation and $5\sigma$ after 2 years.
\end{abstract}

\maketitle
%%%%%%%%%%%%%%%%%%%%%%%%%%%%%%%%%%%%%%%%%%%%%%%%%%%%%%%%%%%%%%%%%%%%%%%%%%%%%%%%%%%%%%%%%%%%%%%%%%%%
\section{Introduction}
%%%%%%%%%%%%%%%%%%%%%%%%%%%%%%%%%%%%%%%%%%%%%%%%%%%%%%%%%%%%%%%%%%%%%%%%%%%%%%%%%%%%%%%%%%%%%%%%%%%%

The standard cosmological  model ($\Lambda$CDM) has had tremendous success in describing our Universe and matching its ever-more precise observations, most notably the measurements of the cosmic microwave background (CMB)~\cite{Planck:2018vyg}. Yet, in recent years several tensions between different observables and the inferences from the model have become more prominent~\cite{Perivolaropoulos:2021jda, Abdalla:2022yfr}. Perhaps the most significant of these is the Hubble tension~\cite{Verde:2019ivm}, which is the discrepancy between local measurements of the current expansion rate of the Universe (see Ref.~\cite{Freedman:2023jcz} for a review), $H_0=73.17\pm0.86\;{\rm km}\;{\rm s}^{-1}\invMpc$~\cite{Breuval:2024lsv}, and the value inferred from the CMB, $H_0=67.4\pm0.5\;{\rm km}\;{\rm s}^{-1}\invMpc$~\cite{Planck:2018vyg}. While this tension, which has reached a level of $5\sigma$, raises questions about possible systematics in our cosmological measurements~\cite{Rigault:2014kaa, NearbySupernovaFactory:2018qkd, Blum:2020mgu, Millon:2019slk, Birrer:2020tax, Freedman:2021ahq}, it also hints at the possibility of new physics beyond $\Lambda$CDM.

A myriad of models has been proposed to address the Hubble tension. Interested readers are encouraged to explore the reviews in Refs.~\cite{DiValentino:2021izs, Schoneberg:2021qvd} for further details. One popular model proposed to elevate the value of $H_0$ inferred from the CMB is that of early dark energy (EDE)~\cite{Karwal:2016vyq, Smith:2019ihp, Kamionkowski:2022pkx}. This model introduces an additional field that behaves as dark energy during the early Universe. Before recombination, around matter-radiation equality, the field undergoes a transition, during which its fractional energy density becomes non-negligible. It then dilutes away shortly after recombination. This transitioning reduces the sound horizon and consequently yields a larger value for $H_0$, effectively alleviating the tension~\cite{Poulin:2018cxd}.

While EDE can potentially increase the inferred value of $H_0$ to align with local measurements, it also introduces additional signatures in other cosmological observables, such as the CMB and large-scale structure (LSS). Various analyses, employing different datasets, have been conducted, leading to divergent conclusions about whether EDE can effectively resolve the Hubble tension~\cite{Poulin:2018cxd, Poulin:2021bjr, Smith:2022hwi, Efstathiou:2023fbn, Murgia:2020ryi, Smith:2020rxx, Poulin:2023lkg, McDonough:2023qcu, Goldstein:2023gnw, Hill:2020osr, Hill:2021yec, Qu:2024lpx, Shen:2024hpx}. Therefore, new data from independent observables may provide complementary constraints on this model.

One of EDE's most prominent signatures is in the matter power spectrum, particularly at small scales, e.g., $k\gtrsim 0.1\;\invMpc$. However, such scales are beyond the reach of current CMB experiments (Planck 2018), or LSS surveys (e.g., SDSS~\cite{Reid:2009xm}, DES~\cite{DES:2017qwj}). While direct measurements of the matter distribution on these small scales have not yet been possible, other observables may be influenced by the power at small scales. It has recently been suggested in Refs.~\cite{Libanore:2022ntl, Adi:2023qdf, Cruz:2023rmo}, that line intensity mapping has the potential to access information on small scales. Particularly useful  is the 21cm  hydrogen line~\cite{Madau:1996cs, Furlanetto:2006jb, Pritchard:2011xb}.

The 21cm signal, generated by the emission or absorption of CMB photons with a wavelength corresponding to the hydrogen hyperfine splitting, is expected to play a pivotal role in cosmology due to its potential to unveil the structure and evolution of the Universe at various cosmic epochs~\cite{Pritchard:2011xb, Madau:1996cs, Barkana:2000fd, Bharadwaj:2004it, Furlanetto:2006jb}. 
While a confirmed measurement of the 21cm signal has yet to be made, several upcoming experiments are  targeting it. Some focus on the global 21cm signal (e.g., EDGES~\cite{Monsalve:2019baw}, SARAS~\cite{2021arXiv210401756N}, LEDA~\cite{2018MNRAS.478.4193P}, REACH~\cite{deLeraAcedo:2022kiu}, and PRIzM~\cite{2019JAI.....850004P}), while others aim to measure its spatial fluctuations (e.g., GMRT~\cite{Pal:2020urw}, MWA~\cite{Yoshiura:2021yfx}, LOFAR~\cite{Mertens:2020llj}, PAPER~\cite{Parsons:2013dwa}, HERA~\cite{DeBoer:2016tnn}, and SKA~\cite{Braun:2015B3}).

In this work, we investigate the effects of EDE on the 21cm signal and demonstrate the potential of upcoming surveys, such as the Hydrogen Epoch of Reionization Array (HERA), to observe these effects. We divide our analysis into three main parts. First, we study the effects of EDE on the 21cm signal due to the shift in cosmological parameters and the additional EDE parameters, particularly the fractional energy density of EDE, $\fede$. We find that EDE leaves significant signatures on the 21cm signal due to the enhanced power on small scales of the matter power spectrum, induced by the shift in the standard cosmological parameters relative to $\Lambda$CDM. Additionally, we find that different values of $\fede$ lead to unique signatures in the 21cm power spectrum. We then consider HERA's expected noise and demonstrate its potential to differentiate between the 21cm power spectra corresponding to the EDE and $\Lambda$CDM models. Finally, we assume a fiducial EDE model and calculate HERA's expected sensitivity to $\fede$ as a function of observation time. These results
%, summarized in Figure~\ref{fig:sensitivity}, 
demonstrate HERA's potential to provide additional constraining power on  EDE  in the coming years. Furthermore, while current constraints on EDE are based on observations at either very low redshifts ($z\lesssim 2$; e.g., SNIa~\cite{Riess:2021jrx}) or extremely high redshifts ($z\sim 1100$; e.g., CMB~\cite{Planck:2018vyg}), the 21cm signal is expected to provide additional constraints from intermediate redshifts, thereby complementing the existing ones.

The rest of this paper is organized as follows: In Section~\ref{sec:formalism}, we briefly review the formalism for the EDE model and the 21cm signal. Section~\ref{sec:analysis} describes the methodology and assumptions of our analysis, while in Section~\ref{sec:results} we present and discuss our results. We conclude in Section~\ref{sec:conclusions}.

\begin{figure*}[t!]
    \centering
    \includegraphics[width=\textwidth]{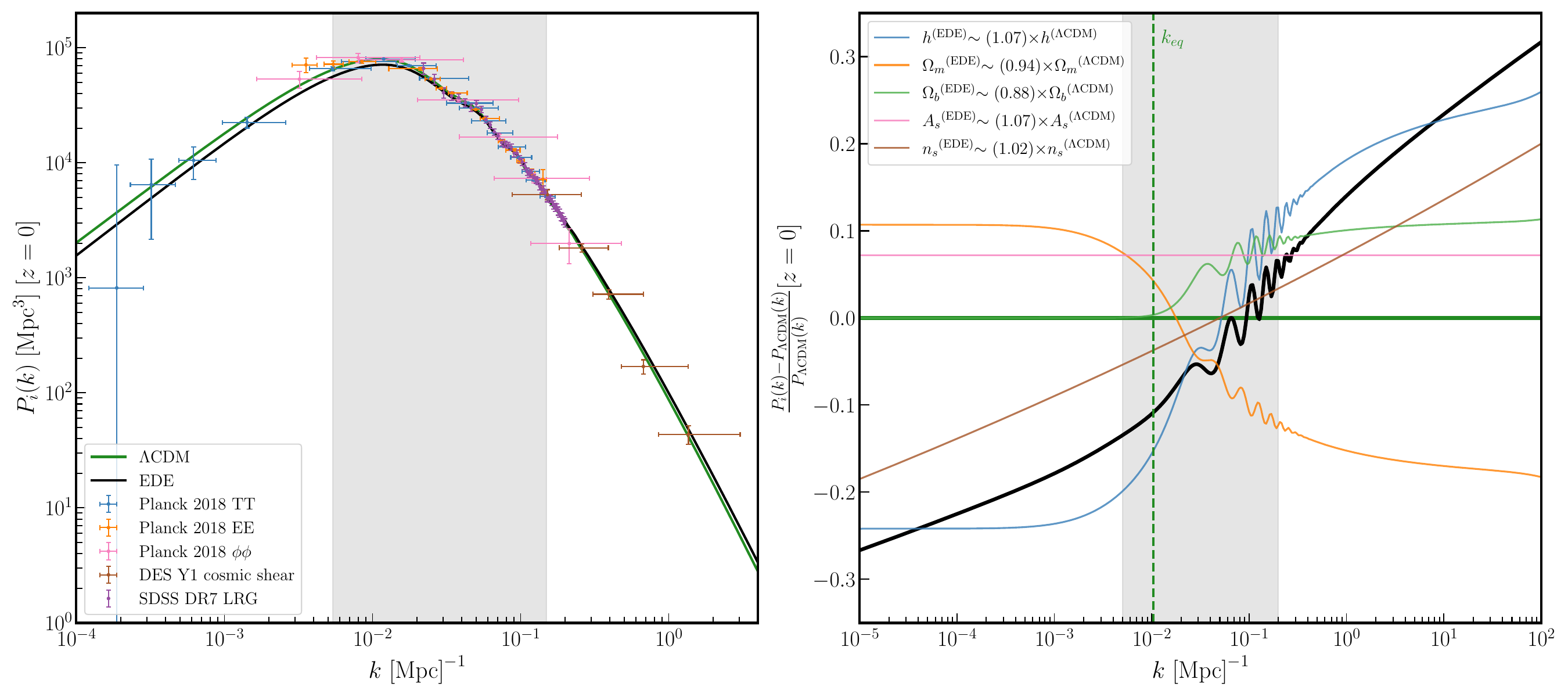}
    \caption{The matter power spectrum. The black lines show the power spectrum for the EDE fiducial model, as defined in Section~\ref{subsec:cosmoparams}, and the green lines represent the power spectrum for the $\Lambda$CDM model, using best-fit values from \textit{Planck 2018}\cite{Planck:2018vyg}.
    Left panel: Error bars correspond to CMB and LSS measurements. The shaded region indicates the scale range where the $P(k)$ error is less than $5\%$.
    Right panel: Residuals relative to $\Lambda$CDM. The power spectrum of the EDE model shows a tilt compared to $\Lambda$CDM due to shifts in standard cosmological parameters (e.g., $n_s$). Additionally, we illustrate the contribution of each cosmological parameter as it shifts from the best-fit $\Lambda$CDM value to that of the EDE fiducial model (relative differences are noted in the legend). For example, one can observe how the decrease in $\Omega_m$ counters the increase in $h$.
    The dashed line denotes the horizon scale at matter-radiation equality.
    }
    \label{fig:mPk}
\end{figure*}

\begin{figure}
    \centering
    \includegraphics[width=\columnwidth]{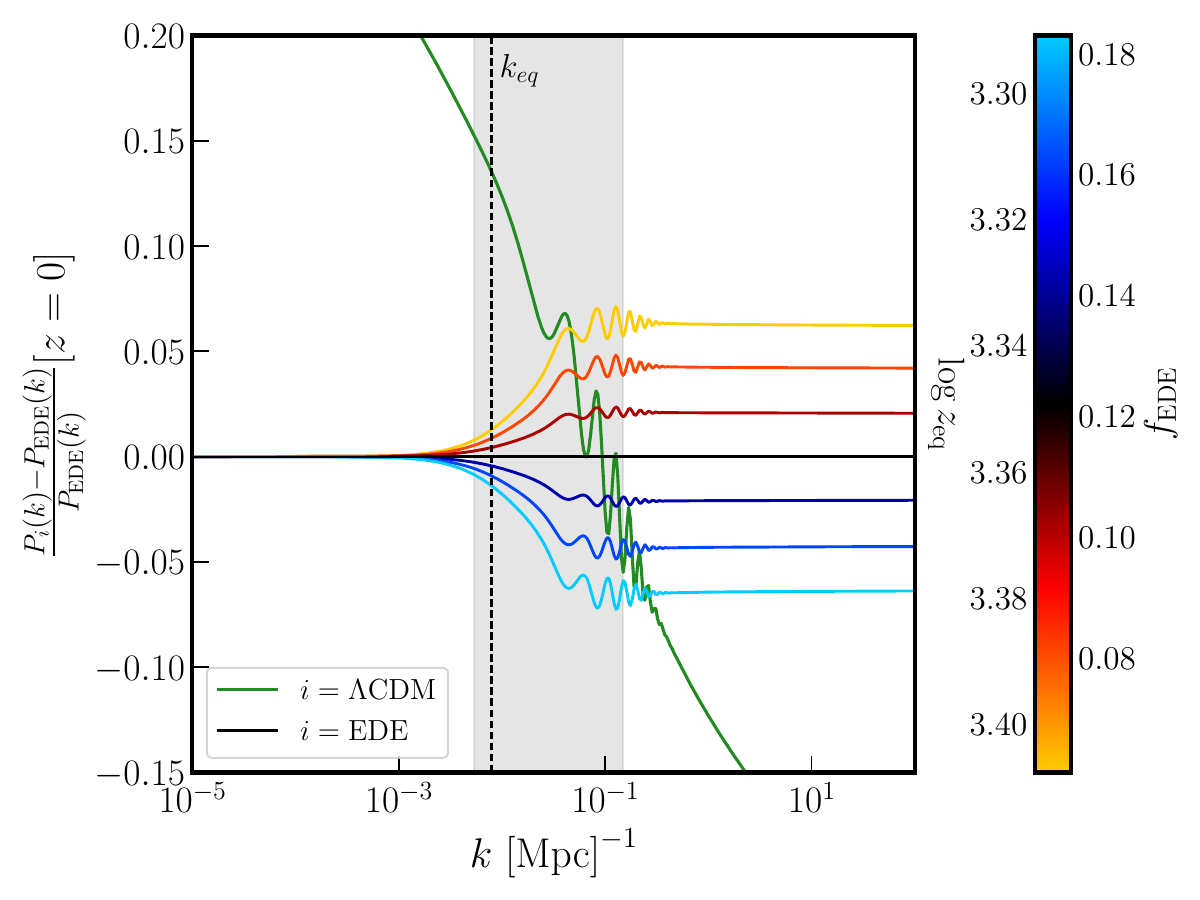}
    \caption{The residuals of the matter power spectrum in the left panel of Figure~\ref{fig:mPk} relative to the EDE fiducial model (black line). The multicolored lines represent different values of $\fede$ around the fiducial value. The dashed line indicates the horizon scale at matter-radiation equality for the fiducial EDE scenario. Additionally, we provide the corresponding values for $z_{\rm eq}$ to illustrate its impact on scales below $k_{\rm eq}$ (with variations in $k_{\rm eq}$ being very small). The shaded region corresponds to the same as in Figure~\ref{fig:mPk}. We emphasize that while the fiducial EDE model enhances power on small scales relative to $\Lambda$CDM due to shifts in standard cosmological parameters, increasing $\fede$ results in power suppression on these scales.}
    \label{fig:mPk_fEDE_res}
\end{figure}

%%%%%%%%%%%%%%%%%%%%%%%%%%%%%%%%%%%%%%%%%%%%%%%%%%%%%%%%%%%%%%%%%%%%%%%%%%%%%%%%%%%%%%%%%%%%%%%%%%%%
\section{Formalism}\label{sec:formalism}
%%%%%%%%%%%%%%%%%%%%%%%%%%%%%%%%%%%%%%%%%%%%%%%%%%%%%%%%%%%%%%%%%%%%%%%%%%%%%%%%%%%%%%%%%%%%%%%%%%%%

Below, we offer a concise overview of the formalism employed in the EDE model and the 21cm signal. For comprehensive reviews of the theory, we refer the reader to Refs.~\cite{Poulin:2018dzj, Smith:2019ihp} and~\cite{Pritchard:2011xb, Furlanetto:2006jb}, respectively

%%%%%%%%%%%%%%%%%%%%%%%%%%%%%%%%%%%%%%%%%%%
\subsection{EDE}\label{subsec:EDE_formalism}
%%%%%%%%%%%%%%%%%%%%%%%%%%%%%%%%%%%%%%%%%%%
We consider a realization of EDE in the form of a minimally coupled scalar field, with the following Klein-Gordon equation of motion
\begin{equation}\label{eq:KGeq}
    \ddot{\phi}+3H\dot{\phi}+V_{n,\phi}=0,
\end{equation}
where the dot denotes a derivative with respect to cosmic time, $H$ is the Hubble parameter, and $V_{n,\phi}\equiv dV_n/d\phi$. The potential of the field is given by
\begin{equation}
    V_n (\phi) = m^2 f^2 \left[ 1 - \cos\left(\phi/f\right)\right]^n,
\end{equation}
such that the only parameters of the EDE model are $f$, $m$ and the initial value of the field $\phi_i$. We follow previous analyses (e.g., Refs.~\cite{Poulin:2021bjr, Efstathiou:2023fbn, Ivanov:2020ril}) and set $n=3$. However, it is convenient to characterize the field by its maximum fraction of the total energy density and its corresponding redshift, such that
\begin{equation}
    \fede \equiv \frac{\rho_\phi(z_c)}{\rho_{\rm tot}(z_c)}.
\end{equation}
For a given $\Theta_i\equiv\phi_i/f$ one can always find $f$ and $m$ which correspond to the phenomenological parameters $\fede$ and $z_c$, so that the three parameters of EDE are \{$\fede$, $z_c$, $\Theta_i$\}.
The background dynamics of the scalar field can be described as follows: at early times, Hubble friction dominates Eq.~\eqref{eq:KGeq}, causing the field to remain frozen at its initial value, and an equation of state $-1$ (akin to dark energy), with a sub-dominant energy density. Only when the Hubble parameter drops below a critical value, dependent on the mass of the field, does the field become dynamic and evolves toward the minimum of its potential. Subsequently, the field oscillates at the bottom of its potential, leading to a dilution of its energy density. For further details, we encourage the reader to explore Ref.~\cite{Poulin:2018dzj}.

The EDE model aims to resolve the Hubble tension by reducing the sound horizon at recombination, as described in Refs.~\cite{Knox:2019rjx, Jedamzik:2020zmd}. During the transition phase of the scalar field, which is set to take place around matter-radiation equality, its energy density peaks and becomes transiently non-negligible. Consequently, the sound horizon at recombination reduces, which leads to the inference of a larger value for $H_0$.

Furthermore, the introduction of EDE into the standard cosmological model affects other observables such as the CMB and the matter power spectrum. Simply put, when analyzing the EDE model with CMB data alone, one does not find any preference for a nonzero $\fede$~\cite{Hill:2020osr, Qu:2024lpx}. However, including data of local $H_0$ measurements drives the best-fit value to $\fede>10\%$. To preserve the fit to CMB data, standard cosmological parameters such as the matter density $\Omega_m$ and spectral index $n_s$ shift from their best-fit values for $\Lambda$CDM~\cite{Hill:2020osr, Smith:2019ihp}. This shift in cosmological parameters manifests most prominently in the tilt of the matter power spectrum, enhancing (suppressing) small (large) scales, as visualized by the black line in Figure~\ref{fig:mPk}. While the fit to the CMB data does not worsen dramatically, the resulting enhancement of power in the matter power spectrum at small scales increases the amplitude of density fluctuations, worsening the existing mild tension with LSS data, known as the $S_8$ tension~\cite{Mccarthy:2017yqf}. Nonetheless, even when accounting for LSS data in the analysis of EDE, one finds that the fit to $H_0$ data dominates, driving the best fit to a nonzero value for $\fede$~\cite{Hill:2020osr}.

 EDE contributes variably to both matter and radiation energy densities throughout its evolution, thereby impacting the redshift of matter-radiation equality, denoted by $z_{\rm eq}$. Specifically, increasing $\fede$ within a given set of cosmological parameters leads to a delay in matter-radiation equality, resulting in a lower value for $z_{\rm eq}$. Consequently, sub-horizon modes with wave numbers $k\gtrsim k_{\rm eq}=a_{\rm eq}H\left(z_{\rm eq}\right)$ have less time to grow in a matter-dominated Universe, leading to their suppression relative to scenarios with lower values of $\fede$. This suppression is illustrated by the multicolored lines in Figure~\ref{fig:mPk_fEDE_res}.

%%%%%%%%%%%%%%%%%%%%%%%%%%%%%%%%%%%%%%%%%%%
\subsection{The 21cm signal}\label{subsec:21cm_formalism}
%%%%%%%%%%%%%%%%%%%%%%%%%%%%%%%%%%%%%%%%%%%
The cosmological 21cm signal, arising from the transition between the triplet and singlet states of the hyperfine splitting in neutral hydrogen, serves as a potent probe for understanding astrophysics and cosmology during the era between recombination and reionization. Throughout this period, a neutral hydrogen atom can be excited to the triplet state by absorbing a CMB photon, colliding with gas particles, and through Ly$\alpha$ pumping~\cite{1952AJ.....57R..31W,1958PIRE...46..240F}.
The 21cm signal is quantified through the differential brightness temperature
\begin{equation}
    T_{21} = \frac{T_s - T_\gamma}{1+z}\left( 1 - e^{-\tau_{21}} \right),
\end{equation}
where $T_\gamma$ is the temperature of the background thermal radiation, typically assumed to be the CMB temperature, and $\tau_{21}$ represents the 21cm optical depth~\cite{Furlanetto:2006jb, Pritchard:2008da}. The spin temperature, $T_s$, quantifies the relative occupation numbers of atoms in triplet and singlet states within the intergalactic medium (IGM), following the relation
\begin{equation}
    T_s^{-1} = \frac{x_\gamma T_\gamma^{-1} + x_c T_k^{-1} + x_\alpha T_c^{-1} }{x_\gamma + x_c + x_\alpha}.
\end{equation}
Here $x_i$, $i=\gamma,c,\alpha$ correspond to the radiation, collision, and Ly$\alpha$ coupling coefficients, respectively, and $T_k$ and $T_c\approx T_k$~\cite{1959ApJ...129..536F} denote the kinetic and color temperatures.

The global 21cm signal, denoted as $\bar{T}_{21}(z)$, is the measured sky-averaged spin temperature against that of the CMB. Post-recombination ($z\sim1100$), the Universe predominantly consists of neutral hydrogen gas, densely packed and prone to frequent collisions. These collisions couple the spin temperature to the kinetic temperature of the gas, with the collision coupling coefficient trumping the radiation coupling coefficient due to the high collision rate. Consequently, the spin temperature follows the adiabatic cooling of the gas, $T_s\approx T_k\propto (1+z)^2$, yielding an absorption feature in the 21cm signal. As the Universe expands, the collision rate drops, tilting the balance towards CMB dominance. Consequently, the absorption signal gradually wanes. This dynamic persists until the emergence of the first stars. These sources emit ultraviolet (UV) radiation, amplifying the Ly$\alpha$ coupling~\cite{Hirata:2005mz}, $x_\alpha$, to a level where the spin temperature is driven again to the kinetic temperature ($T_c\approx T_k$). This strengthened coupling induces a pronounced absorption feature in the 21cm signal. Subsequently, UV and X-ray emissions from stars and their remnants ionize and heat the IGM~\cite{Ciardi:2009zd}, transitioning the 21cm signal from absorption to emission and finally vanquishing  it as the IGM  fully (re)ionizes.

The processes mentioned above also take place at the perturbative level, following the density perturbations, so that the 21cm brightness temperature perturbations are $\delta T_{21}(\bs{x},z)\equiv T_{21}(\bs{x},z)-\bar{T}_{21}(z)$. The statistics of these anisotropies are quantified to  first order via the reduced 21cm dimensionless power spectrum, which is defined as
\begin{equation}
    \Delta_{21}^2(k,z) = \frac{k^3 P_{21}(k,z)}{2\pi^2},
\end{equation}
where $P_{21}(k,z)$ is the angle-averaged Fourier transform of the two-point function $\langle \delta T_{21}(\bs{x},z),\delta T_{21}(\bs{x}',z)\rangle$.

\begin{figure*}
    \centering
    \includegraphics[width=\textwidth]{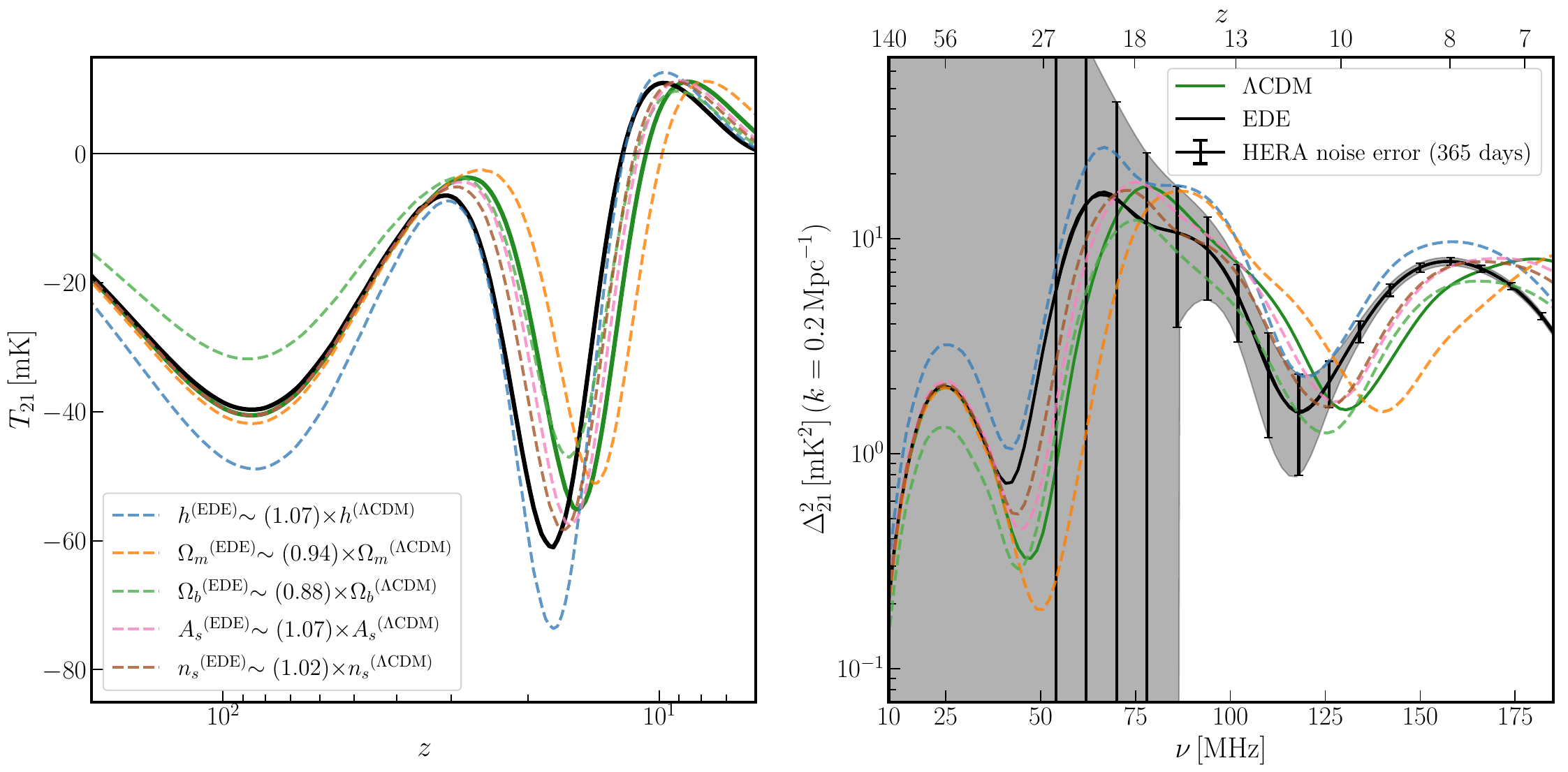}
    \caption{The 21cm signal as a function of redshift. Solid green (black) lines denote the signals corresponding to the $\Lambda$CDM (EDE) model. Dashed lines represent the signal for the $\Lambda$CDM model with one of the cosmological parameters shifted to its EDE value, corresponding to the matter power spectra in Figure~\ref{fig:mPk}. Left panel: the global 21cm signal as a function of redshift. Right panel: the 21cm power spectrum at $k=0.2\,\invMpc$. The error bars correspond to HERA's expected uncertainty after one year of observation.}
    \label{fig:21cm_ofz_cosmo}
\end{figure*}

After cosmic dawn, when $x_\alpha$ becomes dominant, and throughout subsequent epochs, the 21cm signal is shaped by the formation and evolution of astrophysical sources interacting with the IGM. While the abundance and evolution of these sources heavily rely on astrophysical modeling and parameterization, they are ultimately sourced by density fluctuations. Specifically, the star formation rate (SFR) depends on the halo mass function, which itself is derived from the matter power spectrum~\cite{Munoz:2021psm, Dekel:2023ddd}. For example, an increase in power at small scales enhances the abundance of DM halos, forming more galaxies. This, in turn, results in a higher SFR, advancing the epochs of cosmic dawn and reionization to higher redshifts. Conversely, a decrease in power at small scales would have the opposite effect. This effect is demonstrated for the EDE model in Figure~\ref{fig:21cm_ofz_cosmo}, where the 21cm global signal for the fiducial EDE model appears shifted to higher redshifts compared to $\Lambda$CDM due to the enhancement of small scales in the matter power spectrum. The same effect is illustrated for variations of $\fede$, which result in corresponding shifts of the signal (Figure~\ref{fig:21cm_ofz_fEDE}). Furthermore, in addition to the shift in the global signal, the 21cm power spectrum will change due to both the scale dependence of the SFR and the gas density distribution. This property of the 21cm signal makes it a powerful probe of models that affect the matter power spectrum, especially at small scales, such as EDE.

%%%%%%%%%%%%%%%%%%%%%%%%%%%%%%%%%%%%%%%%%%%%%%%%%%%%%%%%%%%%%%%%%%%%%%%%%%%%%%%%%%%%%%%%%%%%%%%%%%%%
\section{Analysis}\label{sec:analysis}
%%%%%%%%%%%%%%%%%%%%%%%%%%%%%%%%%%%%%%%%%%%%%%%%%%%%%%%%%%%%%%%%%%%%%%%%%%%%%%%%%%%%%%%%%%%%%%%%%%%%

Our analysis aims to explore the effects of EDE on the 21cm signal, HERA's potential to differentiate the 21cm power spectra corresponding to EDE and $\Lambda$CDM, and forecast HERA's sensitivity to $\fede$. Our methodology for simulating the 21cm signal and HERA's noise involves the following steps.
First, we utilize the public code \code{AxiCLASS}\footnote{\url{https://github.com/PoulinV/AxiCLASS}}~\cite{Smith:2019ihp, Poulin:2018dzj, Poulin:2018cxd, Murgia:2020ryi}, which incorporates the physics of axion-like fields into the Cosmic Linear Anisotropy Solving System code, \code{CLASS}~\cite{Blas:2011rf}, to simulate the evolution of linear perturbations in the Universe and compute the matter power spectrum. Subsequently, we use the output of \code{AxiCLASS} as the initial conditions for a modified version\footnote{We have made the necessary modifications to replace the native integration of \code{CLASS} with one of \code{AxiCLASS}.} of the semi-numerical public code \code{21cmFirstCLASS}\footnote{\url{https://github.com/jordanflitter/21cmFirstCLASS}}\cite{Flitter:2023mjj, Flitter:2023rzv}, built upon \code{21cmFAST}~\cite{Mesinger:2010ne, Munoz:2021psm}, for simulating the inhomogeneities of the 21cm signal. Finally, we use the public code \code{21cmSense}\footnote{\url{https://github.com/rasg-affiliates/21cmSense}}~\cite{2013AJ....145...65P,2014ApJ...782...66P} for calculating HERA's expected uncertainty in measuring the 21cm power spectrum.

To evaluate HERA's sensitivity to EDE, we use the resulting 21cm power spectrum and its uncertainties to calculate the 21cm Fisher matrix, from which we extract the marginalized uncertainty on $\fede$.

%%%%%%%%%%%%%%%%%%%%%%%%%%%%%%%%%%%%%%%%%%%
\subsection{\code{21cmFirstCLASS}}
\label{subsec:21cmFC}
%%%%%%%%%%%%%%%%%%%%%%%%%%%%%%%%%%%%%%%%%%%
As the name suggests, \code{21cmFirstCLASS} integrates \code{21cmFAST} and \code{CLASS}. By doing so, the code computes the initial conditions (e.g., the matter density and relative velocity transfer functions, the growth function, and the background quantities such as the kinetic temperature and ionization fraction) and then evolves the density fluctuations from recombination to cosmic dawn.
% Additionally, it evolves the baryon density fluctuations instead of CDM, such that $\delta_b\propto\mathcal{D}_b(k,z)$, where
% \begin{equation}
% \mathcal{D}_b(k,z)\equiv \frac{\mathcal{T}_b(k,z)}{\mathcal{T}_c(k,z)}D(z)
% \end{equation}
% is the scale-dependent growth factor, $\mathcal{T}_i$ denotes the transfer functions, and $D(z)$ is the scale-independent growth factor. This correction is important mostly during the dark ages, since by the time of cosmic dawn, the baryons gravitate towards the CDM, i.e., $\delta_b\propto\delta_c\propto D(z)$.

This new code allows for the consistent calculation and study of the 21cm signal in various cosmologies within and beyond $\Lambda$CDM, as demonstrated in this work. For further details, we refer the reader to Refs.~\cite{Flitter:2023mjj, Flitter:2023rzv} and the documentation of \href{https://github.com/jordanflitter/21cmFirstCLASS}{\code{21cmFirstCLASS}}.

In all of our simulations, we used a box with a volume of $(1024\,{\rm Mpc})^3$ and $(256)^3$ voxels. The size of the box allows for sufficient statistics at large scales ($k\gtrsim0.1\,\invMpc$), while the voxel size of $\sim4\,{\rm Mpc}$ provides sufficient resolution for the power spectrum to converge at small scales ($k\sim 0.5\,\invMpc$), aligned with HERA's sensitivity range\footnote{We confirmed this by testing different realizations. For further discussion, we refer the reader to Ref.~\cite{Kaur:2020qsa}.}.
%Additionally, we conducted tests with different realizations to validate the robustness of our results, as summarized in Appendix~\ref{app:realizations_var}.

We note that we did not account for the EDE contribution to the Hubble parameter, $H(z)$, when implementing it in \code{21cmFirstCLASS} since it dilutes faster than radiation after $z_c$ and its contribution at $z<z_{\rm rec}<z_c$ is completely negligible (e.g., Ref.~\cite{Poulin:2018dzj}).

%%%%%%%%%%%%%%%%%%%%%%%%%%%%%%%%%%%%%%%%%%%
\subsection{Cosmological parameters}
\label{subsec:cosmoparams}
%%%%%%%%%%%%%%%%%%%%%%%%%%%%%%%%%%%%%%%%%%%
While EDE allows for accommodating a larger value of $H_0$, it also necessitates shifts in other cosmological parameters to preserve the fit to other data, particularly to CMB~\cite{Smith:2019ihp, Hill:2020osr, Smith:2020rxx}. Therefore, in this analysis, we adopt EDE best-fit parameters from Table~I in Ref.~\cite{Smith:2019ihp} as our fiducial model:
\begin{align}\label{eq:cosmo_params}
    \begin{aligned}
        h &= 0.7219, & 100\omega_b &= 2.253, & \omega_{\rm cdm} &= 0.1306, \\
        10^{9}A_s &= 2.251, & n_s &= 0.9889, & \tau_{\rm reio} &= 0.072, \\
        \log(z_c) &= 3.562, & \fede &= 0.122, & \Theta_i &= 2.83.
    \end{aligned}
\end{align}
Here $h$ represents the dimensionless Hubble constant, while $\omega_i=\Omega_i h^2$ denotes the dimensionless energy density, with $i=b,\rm{cdm}$ corresponding to baryons and cold dark matter (CDM), respectively. Additionally, $A_s$ denotes the amplitude of scalar fluctuations, $n_s$ refers to the scalar spectral index, and $\tau_{\rm reio}$ represents the optical depth to reionization. The parameters $z_c$, $\fede$, and $\Theta_i$ specify the EDE model and are defined in Section~\ref{subsec:EDE_formalism}. Additionally, we adopt the Planck collaboration convention and model free-streaming neutrinos as two massless species and one massive with $m_\nu = 0.06$~eV. These parameters are used for both the computation of the initial conditions using \code{AxiCLASS} and the simulation of the 21cm inhomogeneities with \code{21cmFirstCLASS}.

%%%%%%%%%%%%%%%%%%%%%%%%%%%%%%%%%%%%%%%%%%%
\subsection{Astrophysical parameters}
\label{sec:astro_params}
%%%%%%%%%%%%%%%%%%%%%%%%%%%%%%%%%%%%%%%%%%%
In addition to the cosmological parameters, which determine the background evolution of the Universe and the density fluctuations, \code{21cmFirstCLASS} requires a set of astrophysical parameters that govern star formation and evolution. The code specifically considers galaxies undergoing atomic-cooling and molecular-cooling processes, which are associated with Population II (pop-II) and Population III (pop-III) stars. These galaxies are defined to inhabit halos of different masses and are characterized by their parameters to represent the efficiency of star formation.
We adopt the values for these astrophysical parameters from Table 1 (EOS2021) of Ref.~\cite{Munoz:2021psm}. It is worth noting that unlike the cosmological parameters, which are constrained by multiple observations including the CMB, the astrophysical parameters are subject to large uncertainties due to the scarcity of observations during reionization and cosmic dawn.

\begin{figure*}
    \centering
    \includegraphics[width=\textwidth]{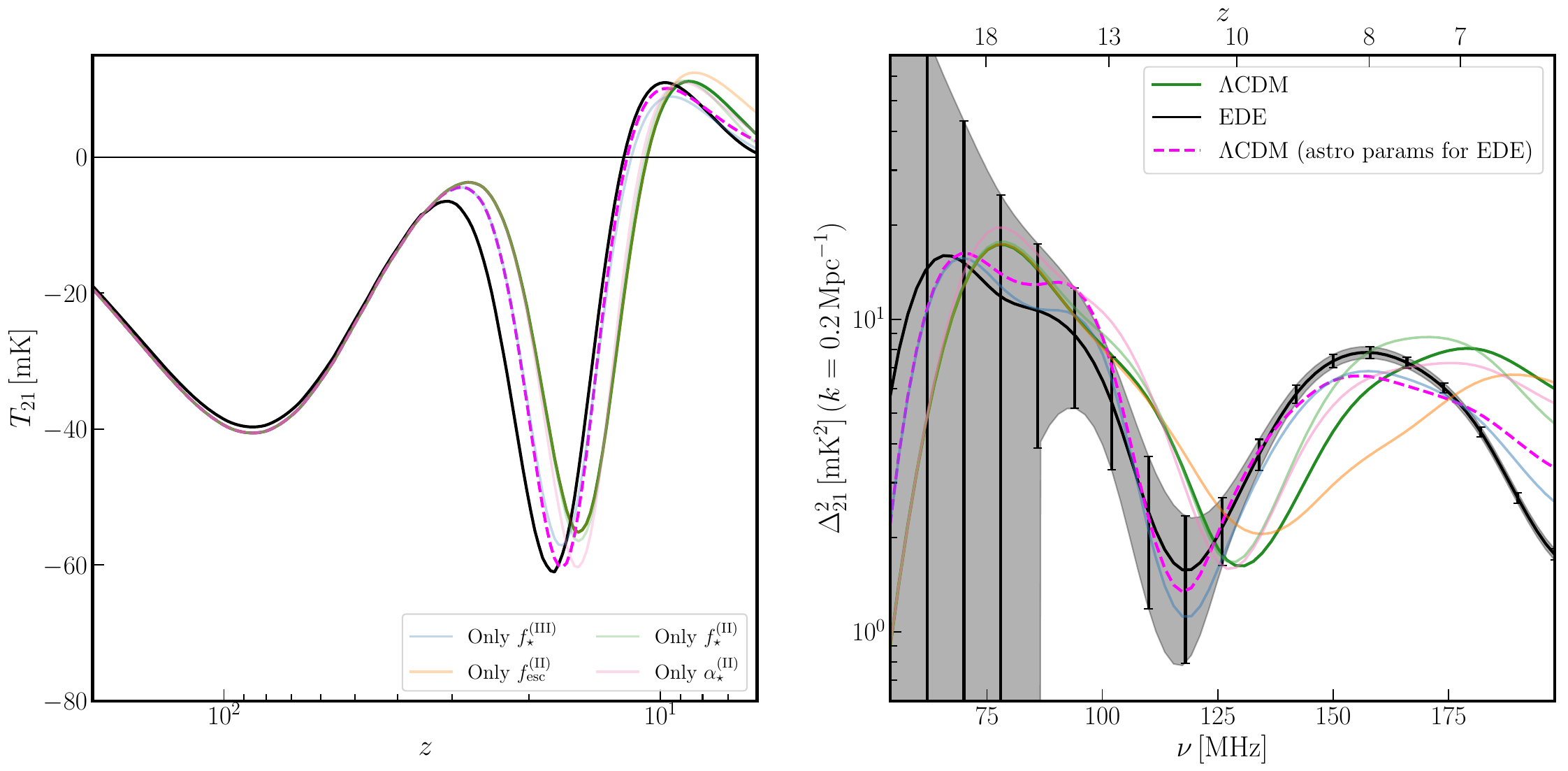}
    \caption{The 21cm signal as a function of redshift. Solid green (black) lines denote the signals corresponding to the $\Lambda$CDM (EDE) model, similar to Figure~\ref{fig:21cm_ofz_cosmo}. Additional solid colored lines depict variations in each of the key astrophysical parameters intended to match the signal characteristics of the EDE model by varying the astrophysical parameters while imposing $\Lambda$CDM's best-fit cosmological parameters. The dashed magenta line corresponds to the combined effect of all parameters, illustrating their collective impact on mimicking the EDE model's signal.}
    \label{fig:21cm_ofz_best_astrofit}
\end{figure*}

\begin{figure}
    \centering
    \includegraphics[width=\columnwidth]{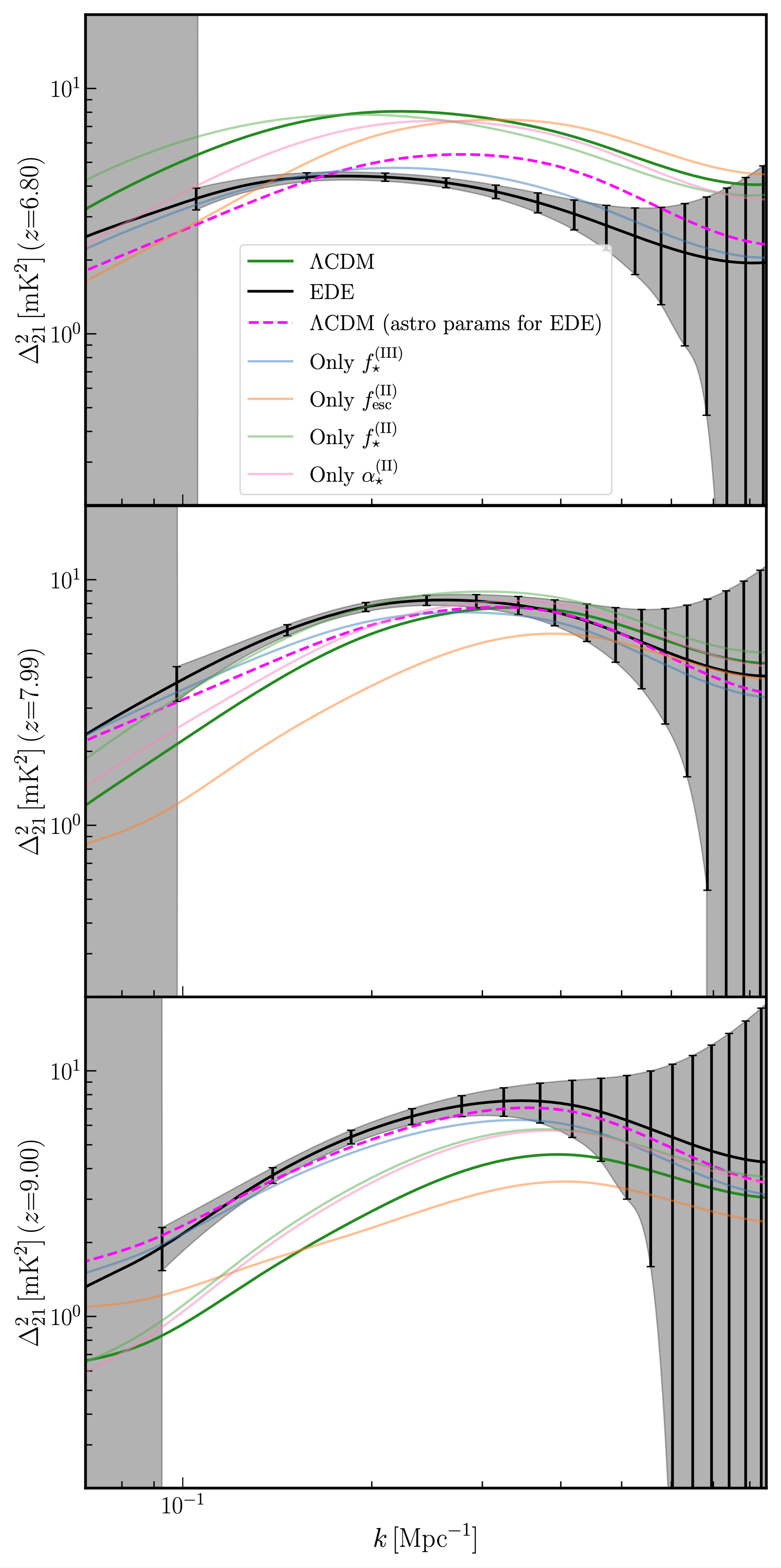}
    \caption{The 21cm power spectrum is plotted as a function of wavenumber for selected redshifts. The lines correspond to those in Figure~\ref{fig:21cm_ofz_best_astrofit}, illustrating the scale dependence of different signals. This complements the right panel of Figure~\ref{fig:21cm_ofz_best_astrofit}.}
    \label{fig:21cmPS_HERA_of_k_astrofit}
\end{figure}

%%%%%%%%%%%%%%%%%%%%%%%%%%%%%%%%%%%%%%%%%%%
\subsection{Noise configuration}
\label{subsec:noise}
%%%%%%%%%%%%%%%%%%%%%%%%%%%%%%%%%%%%%%%%%%%
To determine the uncertainty in the 21cm power spectrum, denoted by $\delta\Delta_{21} (k, z)$, we utilize \code{21cmSense} to simulate the expected noise observed by HERA. HERA's final configuration (\textit{full} HERA) consists of 331 antennae, each 14 meters wide, arranged in a hexagonal pattern with 11 antennae at its base. Operating between 50 MHz ($z = 27.4$) and 225 MHz ($z = 5.3$) with an 8 MHz bandwidth, HERA's analysis focuses on 19 redshift bins, excluding those below $z = 6$ due to uncertainty in reionization details. Each frequency band comprises 82 channels, totaling 1024 channels across 100 MHz.

We assume HERA operates for six hours per night and vary the number of observation days. Additionally, we set the receiver temperature at 100 K and the sky temperature following $T_{\rm sky}(\nu) = 60 K (\nu/300 {\rm MHz})^{-2.55}$. Moreover, we consider both ``moderate" and ``pessimistic" foreground scenarios, for which all baselines are added coherently and incoherently, respectively. Whereas the value of the wedge super-horizon buffer is assumed to extend to $0.1h\invMpc$ beyond the horizon wedge limit, following Refs.~\cite{Flitter:2023mjj, Liu:2019awk, Morales:2012kf}.

% Finally, as described in Appendix~\ref{app:realizations_var}, the simulated signal from \code{21cmFirstCLASS} varies with different random seeds. This variability should not be mistaken for the sample variance considered in \code{21cmSense}, but rather as an artifact of the simulation code. Ideally, the statistical properties of a simulated volume should stabilize for sufficiently large boxes with high resolution. We address this uncertainty in two ways. First, in all our plots, we add an additional $5\%$ of the signal's value to represent the variance across different random realizations (error bars reflect HERA's noise, while shaded regions include this $5\%$ uncertainty). Second, we average the uncertainties derived from our Fisher analysis over 20 simulations with different random seeds.

Throughout this paper, HERA's noise as represented by the error bars in the plots, corresponds to the ``moderate" foreground scenario after one year of observation.

%%%%%%%%%%%%%%%%%%%%%%%%%%%%%%%%%%%%%%%%%%%
\subsection{Fisher analysis}
\label{subsec:fisher}
%%%%%%%%%%%%%%%%%%%%%%%%%%%%%%%%%%%%%%%%%%%
We forecast HERA's sensitivity to EDE's fractional energy density, $\fede$, by employing a Fisher analysis, following the formalism in Refs.~\cite{Jungman:1995bz, Mason:2022obt} to construct the Fisher matrix for the 21cm power spectrum,
\begin{equation}
    F_{i,j}^{\rm 21cm} = \sum_{k,z} \left[ \delta\Delta_{21}^2(k,z) \right]^{-2}\frac{\partial \Delta_{21}^2(k,z)}{\partial\theta_i} \frac{\partial \Delta_{21}^2(k,z)}{\partial\theta_j},
\end{equation}
where $\theta_i$ denotes the cosmological and astrophysical parameters we vary in our analysis\footnote{We ensure that parameters defined logarithmically are varied linearly.},
\begin{align}
    \theta_i\in & \{h,\, \omega_b,\, \omega_{\rm cdm},\, A_s,\, n_s,\,\log(z_c),\, \fede,\, \Theta_i,\nonumber\\
    & L_X^\mathrm{(II/III)},\, f_*^\mathrm{(II/III)},\, f_{\rm esc}^\mathrm{(II/III)},\, \alpha_*^\mathrm{(II/III)},\nonumber\\
    & A_{\rm LW},\, A_{v_{\rm cb}}\}.
\end{align}
Here $L_X^\mathrm{(II/III)}$ are the X-ray luminosity of pop-II/III stars (normalized by the SFR in units of $\rm erg \; s^{-1} M_\odot^{-1} yr$), $f_*^\mathrm{(II/III)}$ are the pop-II/III star formation efficiency, $f_{\rm esc}^\mathrm{(II/III)}$ are the fraction of photons that were produced by pop-II/III stars and managed to escape the host galaxy and ionize the IGM, $\alpha_*^\mathrm{(II/III)}$ are the pop-II/III star formation efficiency power law index, whereas $A_{\rm LW}$ and $A_{v_{\rm cb}}$ are the amplitudes of the Lyman-Werner (LW)~\cite{Haiman:1996rc, Bromm:2003vv} and CDM-baryon relative velocity ($v_{\rm cb}$)~\cite{Tseliakhovich:2010bj, Ali-Haimoud:2013hpa} feedbacks on the minimum molecular cooling galaxies mass that can still host stars, respectively.
We exclude the optical depth parameter, $\tau_{\rm reio}$, as it does not affect the matter power spectrum and is calculated within \code{21cmFirstCLASS}, as demonstrated in Ref.~\cite{Shmueli:2023box}. Finally, we evaluate the covariance matrix by inverting the Fisher matrix.

\begin{figure*}
    \centering
    \includegraphics[width=\textwidth]{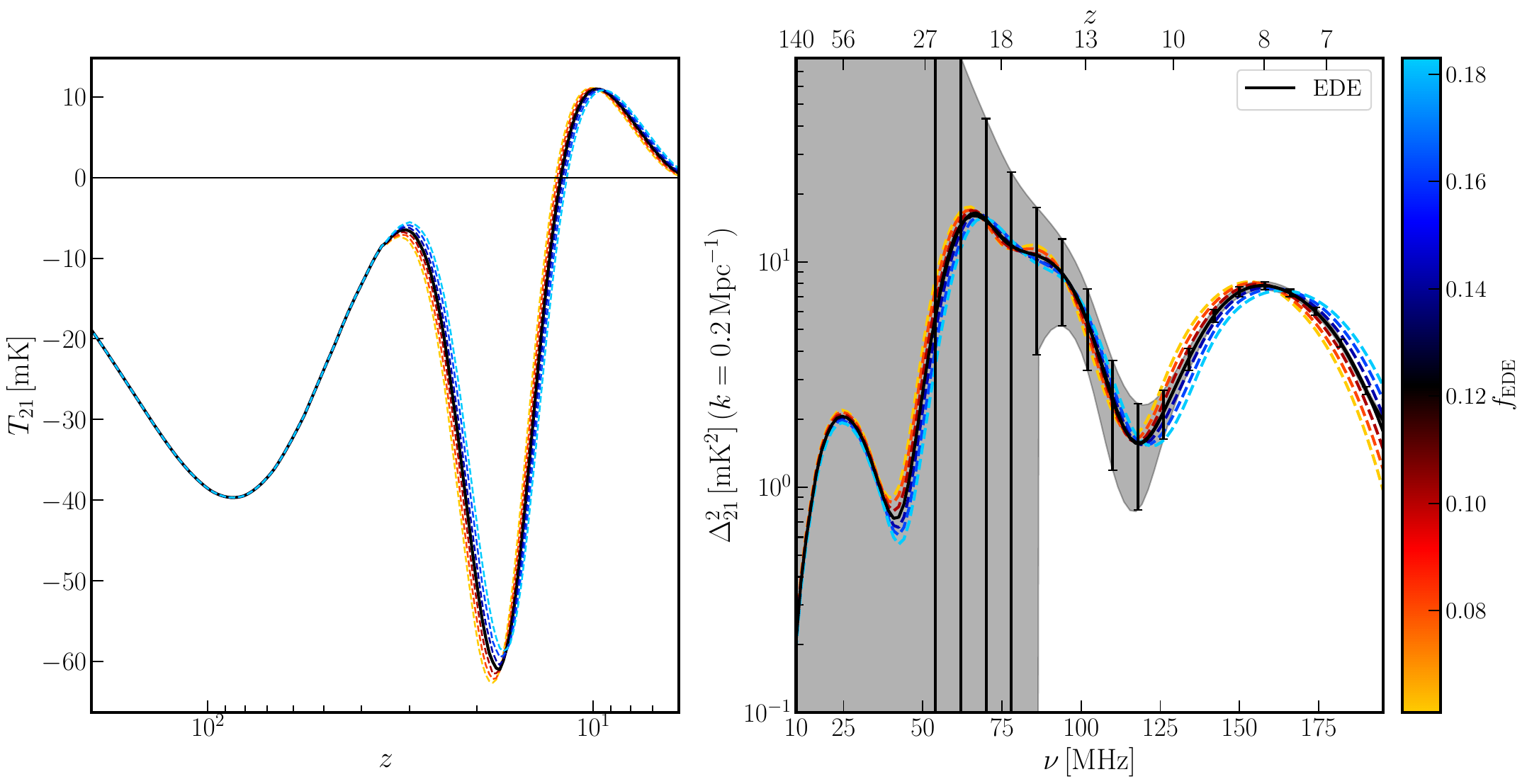}
    \caption{The 21cm signal as a function of redshift. Solid black lines denote the signals corresponding to the EDE model, similar to Figure~\ref{fig:21cm_ofz_cosmo}. Dashed multicolored lines depict variations in $\fede$, corresponding to Figure~\ref{fig:mPk_fEDE_res}. It is evident that the effects of $\fede$ on the global 21cm signal are prominent only after cosmic dawn, whereas its effects on the power spectrum manifest before then due to the impact on density fluctuation.}
    \label{fig:21cm_ofz_fEDE}
\end{figure*}

\begin{figure}
    \centering
    \includegraphics[width=\columnwidth]{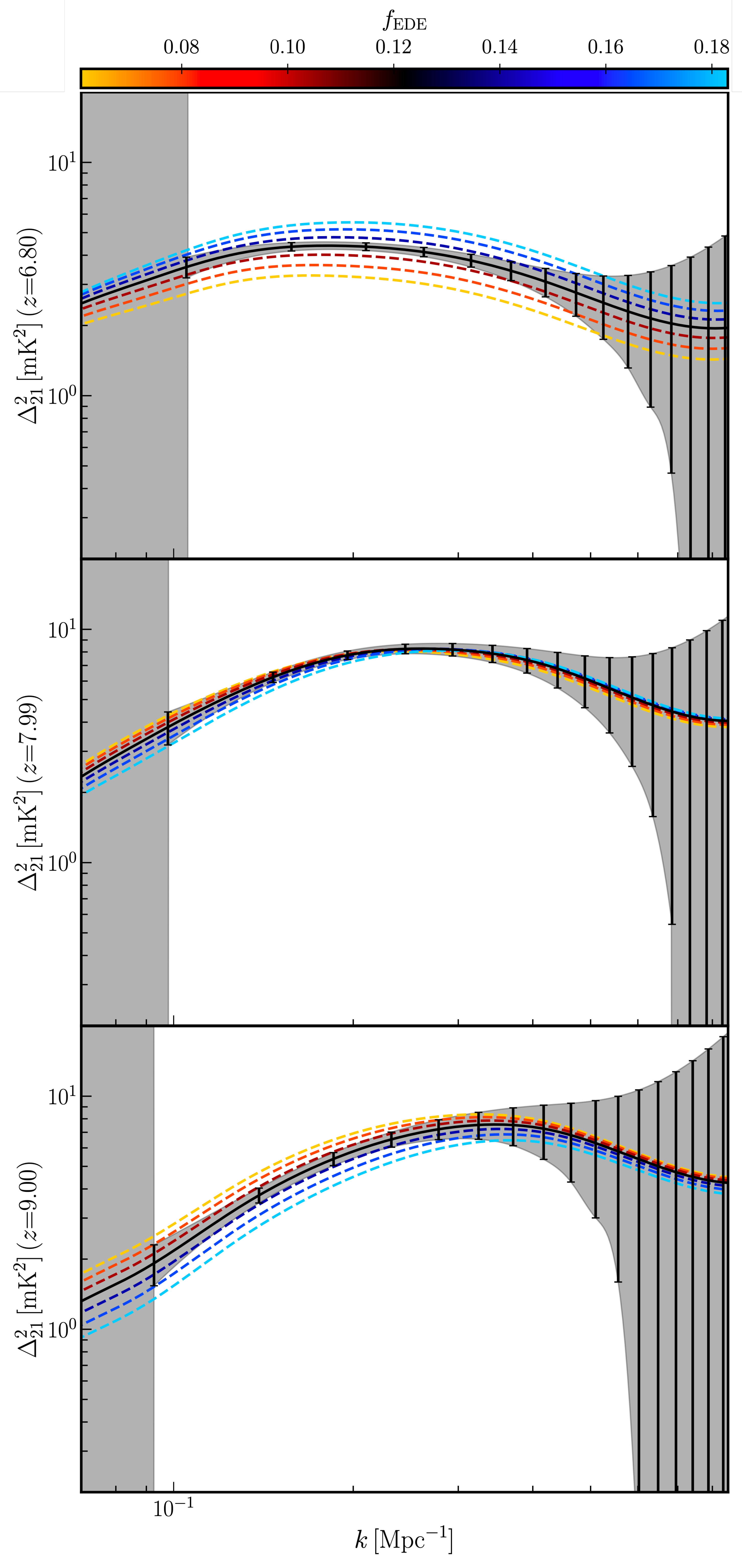}
    \caption{The 21cm power spectrum is plotted as a function of wavenumber for selected redshifts. The lines correspond to those in Figure~\ref{fig:21cm_ofz_fEDE}, illustrating the scale dependence of different signals. This complements the right panel of Figure~\ref{fig:21cm_ofz_fEDE}.}
    \label{fig:21cmPS_HERA_of_k_fEDE}
\end{figure}

%%%%%%%%%%%%%%%%%%%%%%%%%%%%%%%%%%%%%%%%%%%%%%%%%%%%%%%%%%%%%%%%%%%%%%%%%%%%%%%%%%%%%%%%%%%%%%%%%%%%
\section{Results and discussion}
\label{sec:results}
%%%%%%%%%%%%%%%%%%%%%%%%%%%%%%%%%%%%%%%%%%%%%%%%%%%%%%%%%%%%%%%%%%%%%%%%%%%%%%%%%%%%%%%%%%%%%%%%%%%%

%%%%%%%%%%%%%%%%%%%%%%%%%%%%%%%%%%%%%%%%%%%
\subsection{EDE effect on the 21cm signal}
%%%%%%%%%%%%%%%%%%%%%%%%%%%%%%%%%%%%%%%%%%%

The effect of EDE on the 21cm signal is primarily dominated by the shift in the standard cosmological parameters, as mentioned in Section~\ref{subsec:EDE_formalism}. To understand this, it is instructive to first analyze the effects of EDE on the matter power spectrum, shown in Figures~\ref{fig:mPk}-\ref{fig:mPk_fEDE_res}. The overall effect of EDE on the matter power spectrum results in the enhancement of small scales and suppression of larger scales, pivoted around $k\sim0.05$ $\rm Mpc^{-1}$, which originates mostly from the increase in $h$ and $n_s$. Additionally, the reduction in $\Omega_b$ (while $\Omega_m$ is kept fixed) leads to a larger CDM-to-baryon ratio, enhancing the power on scales smaller than $k_{\rm rec}\lesssim k_{\rm eq}$, suppressed by baryon-photon coupling before recombination. A similar effect is observed when considering the lower $\Omega_m$ (while $\Omega_b$ is kept fixed).

The magnitude and scale dependence of the resulting changes in the matter power spectrum due to the shift of each cosmological parameter propagate into both the global signal and power spectrum of the 21cm signal, as shown in Figure~\ref{fig:21cm_ofz_cosmo}. While some cosmological parameters (e.g., $h$, $\Omega_m$, and $\Omega_b$) affect the signal before cosmic dawn, their contributions roughly cancel each other out, resulting in a small suppression of the absorption signal during the dark age ($z\sim250$). Specifically, the increase in $h$ leads to stronger coupling of $T_s$ to $T_k$, increasing the absorption signal, whereas the reduction in $\Omega_b$ has the opposite effect. In general, as expected, we observe an enhancement of all the features of the global signal when $ h$ is increased. Therefore, most of the effects of EDE on the 21cm signal appear after cosmic dawn due to the modifications in the matter power spectrum.

After cosmic dawn ($z \sim 35$), the global signal is affected mostly by the matter power spectrum at small scales, where increased power leads to an earlier onset of cosmic dawn and consequently a shift in the global signal. This is evident from the difference between the curve corresponding to the shift in $\Omega_m$ and those of the other parameters, relative to the best-fit $\Lambda$CDM (solid dark green) in Figure~\ref{fig:21cm_ofz_cosmo}. While these features in the global signal also dominate the evolution of the power spectrum, they occur at different redshifts and scales, providing a powerful means of differentiating and constraining different models.

While the shift in the standard cosmological parameters predominantly determines the shape of the matter power spectrum for the EDE model and thus its effects on the 21cm signal, compared to $\Lambda$CDM, the fractional energy density of EDE, $\fede$, also leaves a unique signature on the matter power spectrum, as described in Section~\ref{subsec:EDE_formalism}. In particular, the increase (decrease) of $\fede$ leads to suppression (enhancement) of scales smaller than $k_{\rm eq}$, as shown in Figure~\ref{fig:mPk_fEDE_res}. Although this feature resembles that of varying $\Omega_b$ (Figure~\ref{fig:mPk}), $\fede$ affects only early physics before recombination, leaving its imprint on the density fluctuations, while baryon density plays a role in the thermal evolution of the neutral hydrogen on the background level as well. This is evident in Figure~\ref{fig:21cm_ofz_cosmo}, where the shift in $\Omega_b$ affects the entire evolution of the global 21cm signal, and in Figure~\ref{fig:21cm_ofz_fEDE}, where variations in $\fede$ affect it only at and from cosmic dawn, once the density fluctuations become non-linear. As expected, the suppression (enhancement) of power at small scales delays (advances) the evolution of the 21cm signal, which, again, affects both the global and the power spectrum of the 21cm signal.

%%%%%%%%%%%%%%%%%%%%%%%%%%%%%%%%%%%%%%%%%%%
\subsection{Differentiating between EDE and $\Lambda$CDM}
%%%%%%%%%%%%%%%%%%%%%%%%%%%%%%%%%%%%%%%%%%%

While Figure~\ref{fig:21cm_ofz_cosmo} suggests that HERA will be able to differentiate between $\Lambda$CDM and EDE best-fit models, it is crucial to remember that this plot assumes a set of astrophysical parameters, many of which are not well constrained (e.g., pop-III parameters), as mentioned in Section~\ref{sec:astro_params}. To check if HERA will be able to distinguish between the two models, one should find the set of astrophysical parameters that yields a signal that mimics the 21cm signal for the EDE's fiducial model while using the $\Lambda$CDM best-fit parameters. Ideally, one should scan the astrophysical parameter space, for example, using an MCMC analysis, to achieve this. However, this is a computationally expensive operation, so we resorted to a less rigorous yet efficient method. Below, we refer to our results in Figures~\ref{fig:21cm_ofz_best_astrofit}-\ref{fig:21cmPS_HERA_of_k_astrofit}.

By considering the imprint of each of the astrophysical parameters on the 21cm signal, we narrowed the number of parameters that affect it in a similar way to the fiducial EDE model down to four main parameters\footnote{We considered the variation of other astrophysical parameters as well but found that the best fit to the EDE scenario required changing only these parameters, at least for the initial coarse scan of the parameter space.}: $f_\star^{\rm III}$, $f_\star^{\rm II}$, $f_{\rm esc}^{\rm III}$, and $\alpha_\star^{\rm II}$. These can be explained as follows.
The earlier epoch at and right after cosmic dawn is dominated by the pop-III parameters. Therefore, to have an earlier onset of galaxy and star formation, as in the EDE scenario, one should adopt a higher efficiency pop-III SFR. This can be achieved by increasing $f_\star^{\rm III}$ (light blue line). However, this will also lead to faster heating and ionization of the IGM. Consequently, the signal will not have enough time to reach the minimum $T_{21}$ before the heating phase, as observed in the EDE scenario, and will also not reach the maximum value right before reionization. To compensate for these effects, a higher pop-II efficiency (light green line), $f_\star^{\rm II}$, and a lower power index for the mass dependence (light red line), $\alpha_\star^{\rm II}$, lead to stronger Ly$\alpha$ coupling, thus producing a stronger absorption feature. Additionally, reducing the ionizing photons escape fraction (light orange line), $f_{\rm esc}^{\rm III}$, prolongs the ionization of the IGM, allowing the signal to reach a higher temperature before it extinguishes.

We scanned over $2\times10^3$ configurations of these four parameters around the fiducial values. We found the one that best fits the signal of the EDE scenario\footnote{We weighted the differences between the signals ($\chi^2$) by HERA's noise for the EDE scenario.}, with $f_\star^{\rm III}=-2.2$, $f_\star^{\rm II}=-1.15$, $f_{\rm esc}^{\rm III}=-1.59$, and $\alpha_\star^{\rm II}=0.2$ (dashed magenta line). Our conclusion is that it is unlikely to identify a set of astrophysical parameters (with the $\Lambda$CDM best-fit cosmological parameters fixed) that will mimic the 21cm signal of the EDE model we consider, even when focusing on the lower redshifts at which HERA is most sensitive. In Figure~\ref{fig:21cmPS_HERA_of_k_astrofit}, we show the scale dependence of the signal for three selected redshifts. One can roughly track the evolution of the power spectrum as power shifts from smaller scales at $z=9$ to larger scales at $z=6.8$, as well as the delay in the overall amplitude for the \textit{astro best fit} and the $\Lambda$CDM scenarios compared to that of EDE.
We note, for completeness, that a more thorough scan of the astrophysical parameter space is necessary to confirm that HERA will be capable of differentiating between EDE and $\Lambda$CDM. Yet, our results provide a strong indication of that capability.

%%%%%%%%%%%%%%%%%%%%%%%%%%%%%%%%%%%%%%%%%%%
\subsection{Sensitivity to $\fede$}
\label{subsec:sensitivity}
%%%%%%%%%%%%%%%%%%%%%%%%%%%%%%%%%%%%%%%%%%%

\begin{figure}
    \centering
    \includegraphics[width=\columnwidth]{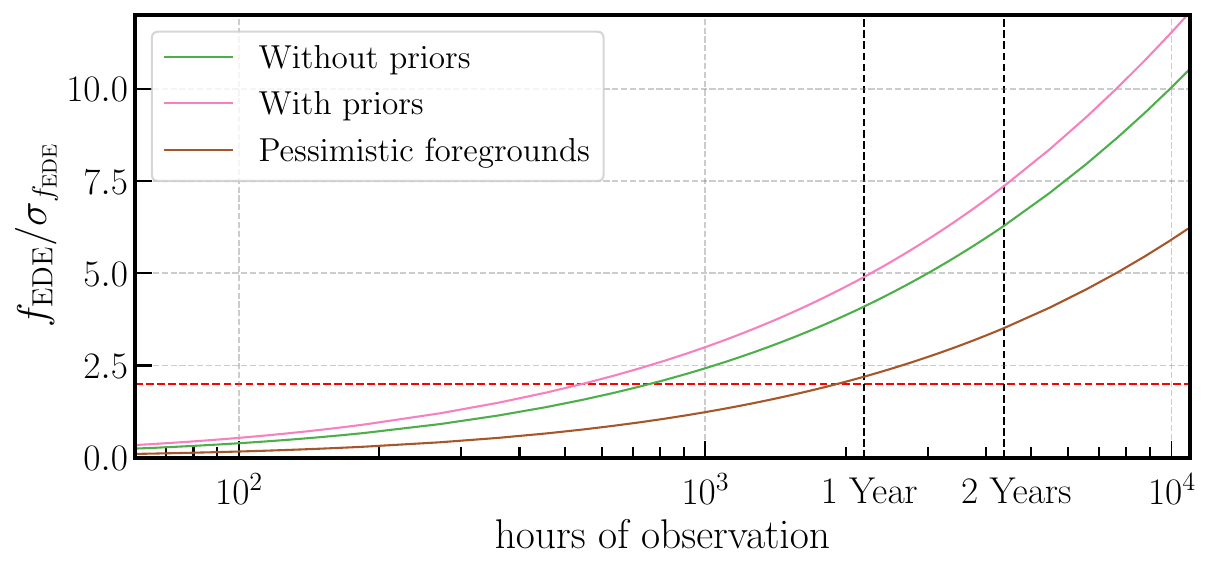}
    \caption{The sensitivity of \textit{full} HERA to $\fede$ as a function of observation time. The green line represents the sensitivity without priors on standard cosmological parameters, the pink line indicates the sensitivity with such priors, and the brown line denotes the sensitivity without priors and with pessimistic foregrounds. To aid interpretation, a red dashed line indicates the sensitivity at $2\sigma$, and black dashed lines correspond to 1 and 2 years of observations (with 6 hours/day).}
    \label{fig:sensitivity}
\end{figure}

Finally, we consider the constraining power of the 21cm power spectrum, expected to be observed by HERA, on the fractional energy density of EDE, $\fede$. We carried out a Fisher analysis, as described in Section~\ref{subsec:fisher}, using the fiducial EDE model detailed in Section~\ref{subsec:cosmoparams} and marginalizing over all other cosmological and astrophysical parameters. Our main results for HERA's sensitivity to $\fede$, expressed as $\fede/\sigma_{\fede}$ (where $\sigma_{\fede}$ is the uncertainty in $\fede$), are shown in Figure~\ref{fig:sensitivity}.

We find that \textit{full} HERA is expected to achieve sensitivity to $\fede$ at a level of $2\sigma$ after approximately 90 days of observations and surpass $5\sigma$ after two years. Notably, incorporating priors on standard cosmological parameters, drawn from the $1\sigma$ uncertainties listed in Table I of Ref.~\cite{Smith:2019ihp}, does not significantly enhance sensitivity. However, under a pessimistic foreground scenario, we observe a substantial drop in sensitivity, with the $2\sigma$ level expected only after 330 days of observation.

Since the sensitivity is dominated by data points where the derivative with respect to $\fede$ is large, some redshift slices contribute more than others, while some may be completely subdominant. This is illustrated in Figure~\ref{fig:21cm_ofz_fEDE}, where the power spectrum lines for different values of $\fede$ overlap. For instance, Figure~\ref{fig:21cmPS_HERA_of_k_fEDE} shows that at redshift $z \approx 8$, the signal is almost invariant to changes in $\fede$.

Additionally, the sensitivity time dependence shown in Figure~\ref{fig:sensitivity} indicates that thermal noise, which is inversely proportional to time (see, for example, Ref.~\cite{2013AJ....145...65P}), dominates over the sample variance error throughout the observed time interval. However, there will be a point where sensitivity saturates as the thermal noise becomes much smaller than the sample variance error.

We note that Fisher analyses are sensitive to derivative instabilities and the chosen fiducial model (i.e., the specific point in parameter space). Therefore, our results should be considered an estimate of HERA's sensitivity to $\fede$ for the given fiducial set of cosmological and astrophysical parameters\footnote{We tested the impact of varying each parameter within a $\sim5\%$ range and found that the results are qualitatively consistent.}. For interested readers,  Figures~\ref{fig:astro_triangle}-\ref{fig:cosmo_triangle} present the covariance plots for the astrophysical and cosmological parameters, including $\fede$.

%%%%%%%%%%%%%%%%%%%%%%%%%%%%%%%%%%%%%%%%%%%%%%%%%%%%%%%%%%%%%%%%%%%%%%%%%%%%%%%%%%%%%%%%%%%%%%%%%%%%
\section{Conclusions}
\label{sec:conclusions}
%%%%%%%%%%%%%%%%%%%%%%%%%%%%%%%%%%%%%%%%%%%%%%%%%%%%%%%%%%%%%%%%%%%%%%%%%%%%%%%%%%%%%%%%%%%%%%%%%%%%

In this work, we investigated the effects of EDE on the 21cm signal and demonstrated the potential of upcoming experiments, such as HERA, to observe these effects. The impact of EDE on small scales in the matter power spectrum influences the number of dark matter halos, which in turn affects the formation and evolution of the first galaxies at redshifts $z \lesssim 30$. These effects shape the 21cm global signal and its anisotropies, making experiments like HERA promising probes for cosmological models such as EDE. Our analysis is divided into three parts: first, we discussed the effects of the EDE model on the 21cm signal; second, we evaluated HERA's potential to differentiate between EDE and $\Lambda$CDM; and finally, we provided a forecast for HERA's sensitivity to the fractional energy density of EDE, $\fede$.

The effects of EDE on the 21cm signal can be attributed to two main contributors: shifts in the standard cosmological parameters and the EDE parameters themselves. We demonstrated that the overall enhancement of small scales in the matter power spectrum within the EDE scenario, compared to $\Lambda$CDM, primarily results from the former. This enhancement leads to both an advancement of the global 21cm signal to earlier times (or higher redshifts) and a distinct morphology of the 21cm power spectrum, as shown in Figure~\ref{fig:21cm_ofz_cosmo}. Additionally, we illustrated that the contribution of the EDE parameters, particularly $\fede$, suppresses sub-horizon scales in the matter power spectrum before matter-radiation equality. Although this effect introduces a degeneracy between $\fede$ and $\Omega_b$ in the matter power spectrum, it does not affect the 21cm signal in the same way. The 21cm signal is affected by $\Omega_b$ through the thermal evolution of hydrogen, leading to a unique signature of $\fede$ that becomes most prominent after cosmic dawn, as shown in Figure~\ref{fig:21cm_ofz_fEDE}. This also manifests in the mild correlation between the two parameters in Figure~\ref{fig:cosmo_triangle}.

To distinguish the 21cm power spectrum in the EDE scenario from $\Lambda$CDM, it is crucial to confirm that no set of astrophysical parameters can mimic the features of the EDE model within the redshifts and scales to which HERA is sensitive. We focused on four astrophysical parameters that showed promise in modifying the power spectrum in the desired direction and scanned their parameter space around the fiducial values to minimize the square difference between the resulting signal and that of the EDE scenario. We found that the signal corresponding to the best-fit setup within the parameter space we covered is still expected to be distinguishable by HERA from that of the EDE scenario, as shown in Figures~\ref{fig:21cm_ofz_best_astrofit}-\ref{fig:21cmPS_HERA_of_k_astrofit}. However, a more comprehensive and efficient analysis is needed to confirm this result, e.g.\ by using a dedicated emulator to explore the parameter space, e.g., Ref.~\cite{Lazare:2023jkg}.

Finally, we provided a forecast for HERA's sensitivity to $\fede$, assuming a fiducial cosmological and astrophysical model. Our results, depicted in Figure~\ref{fig:sensitivity},  indicate that HERA will achieve significant sensitivity to $\fede$ after a relatively short period of observation. For instance, \textit{full} HERA is expected to reach a sensitivity of $2\sigma$ after roughly 90 days and exceed $5\sigma$ after  one year. In contrast, under a more conservative approach considering a pessimistic foreground scenario (see Section~\ref{subsec:noise}), $2\sigma$ sensitivity is expected only after almost a year of observation. These results highlight HERA's potential to probe the matter power spectrum at small scales, particularly those influenced by EDE, well before next-generation experiments, such as CMB-S4 \cite{CMB-S4:2016ple}, become available.

\begin{acknowledgments}

We thank Sarah Libanore for  useful discussions and comments. TA and JF are supported by the Negev Scholarship of the Kreitman School at Ben-Gurion University. EDK acknowledges  joint support from the U.S.-Israel Bi-national Science Foundation (BSF, grant No. 2022743) and  the U.S. National Science Foundation (NSF, grant No. 2307354), as well as support from the ISF-NSFC joint research program (grant No. 3156/23).

\end{acknowledgments}

%%%%%%%%%%%%%%%%%%%%%%%%%%%%%%%%%%%%%%%%%%%%%%%%%%%%%%%%%%%%%%%%%%%%%%%%%%%%%%%%%%%%%%%%%%
%%%%%%%%%%%%%%%%%%%%%%%%%%%%%%%%%%%%%%%%%%%%%%%%%%%%%%%%%%%%%%%%%%%%%%%%%%%%%%%%%%%%%%%%%%
% \newpage
\appendix
\setcounter{footnote}{0}

\newpage

\bibliographystyle{apsrev4-1}
\bibliography{refs}

\begin{figure*}
    \centering
    \includegraphics[width=\textwidth]{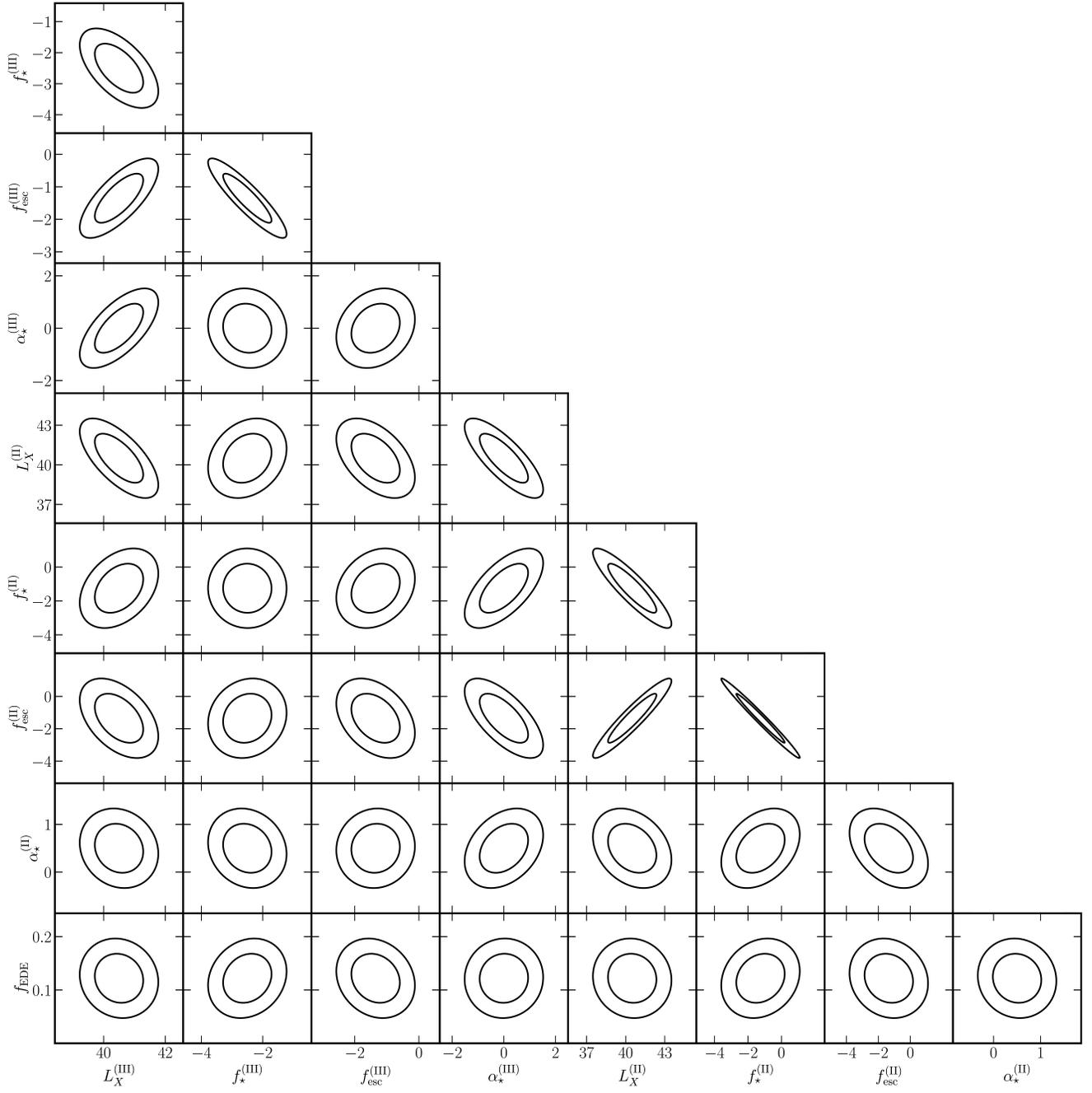}
    \caption{Degeneracies among astrophysical parameters, examined for the 21cm power spectrum with HERA noise after 365 days of observation. The ellipses denote $1\sigma$ and $2\sigma$ confidence levels.}
    \label{fig:astro_triangle}
\end{figure*}

\begin{figure*}
    \centering
    \includegraphics[width=\textwidth]{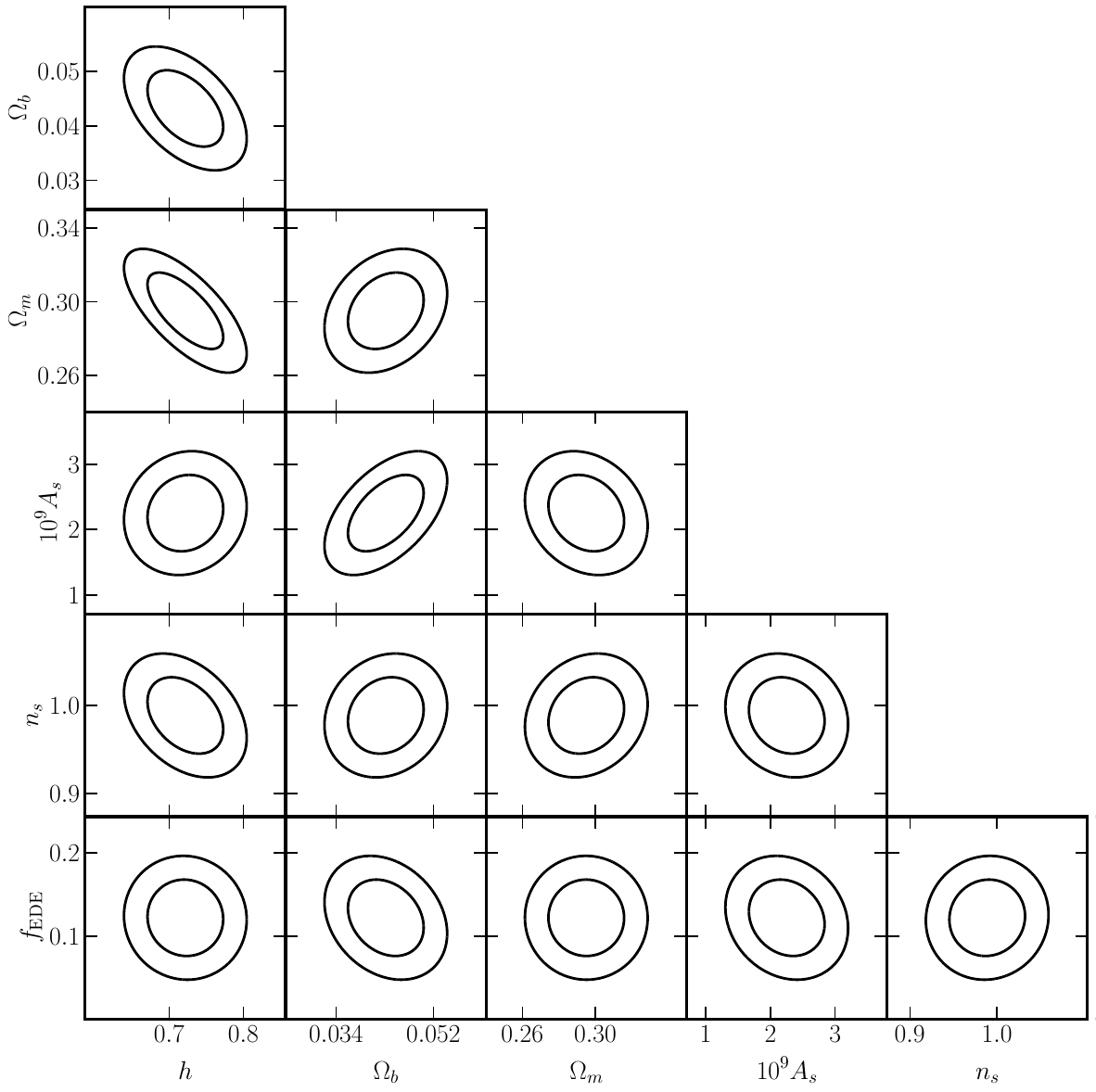}
    \caption{Degeneracies among cosmological parameters, examined for the 21cm power spectrum with HERA noise after 365 days of observation. The ellipses denote $1\sigma$ and $2\sigma$ confidence levels.}
    \label{fig:cosmo_triangle}
\end{figure*}

\end{document}